\documentclass[aps,prd,twocolumn,showpacs,superscriptaddress]{revtex4}  
\usepackage{graphicx}  
\usepackage{dcolumn}   
\usepackage{bm}        
\usepackage{amssymb}   
\usepackage{amsmath,color}
\usepackage{hyperref, url}
\usepackage{float}
\usepackage{adjustbox}
\usepackage{multirow}
\newcommand{\comment}[1]{}

\begin{document}

\widetext
\leftline{Version xx as of \today}


\title{Forward modelling of quasar light curves and the cosmological matter power spectrum on milliparsec scales}

\author{Mansour Karami}\affiliation{Department of Physics and Astronomy, University of Waterloo, 200 University Avenue West, Waterloo, ON, N2L 3G1, Canada}
 \affiliation{Perimeter Institute for Theoretical Physics, 31 Caroline Street North, Waterloo, ON, N2L 2Y5, Canada} 
\author{Niayesh Afshordi} \affiliation{Department of Physics and Astronomy, University of Waterloo, 200 University Avenue West, Waterloo, ON, N2L 3G1, Canada}
 \affiliation{Perimeter Institute for Theoretical Physics, 31 Caroline Street North, Waterloo, ON, N2L 2Y5, Canada}
 \author{Jes\'us Zavala} \affiliation{Center for Astrophysics and Cosmology, Science Institute, University of Iceland, Dunhagi 5, 107 Reykjavik, Iceland}

\date{\today}

\begin{abstract}
  We devise an optimal method to measure the temporal power spectrum of the lensing and intrinsic fluctuations of multiply-imaged strongly lensed quasar light curves, along with the associated time delays.
  The method is based on a Monte-Carlo Markov Chain (MCMC) sampling of a putative gaussian likelihood, and accurately recovers the input properties of simulated light curves, as well as the ``Time Delay Challenge''. We apply this method to constrain the dimensionless cosmological
  (non-linear) matter power spectrum on milliparsec scales (comparable to the size of the solar system), to $\Delta_{\rm NL}^2< 4 \times 10^7$ at $k_{\rm NL} \sim  10^3 {\rm pc}^{-1}$. Using a semi-analytic nonlinear clustering model which is calibrated to simulations, the corresponding constraint on the primordial (linear) scalar power spectrum is ${\cal P}_{\cal R} < 3 \times 10^{-9}$ at $k_{\rm L} \sim$ 3 pc$^{-1}$. This is the strongest constraint on primordial power spectrum at these scales, and is within an order of magnitude from the standard $\Lambda$CDM prediction.   We also report measurements of temporal spectra for intrinsic variabilities of quasar light curves, which can be used to constrain the size of the emitting region in accretion disks. Future cadenced optical imaging surveys, such as LSST, should increase the number of observed strongly lensed quasars by 3 orders of magnitude and significantly improve these measurements, even though improvements in modelling quasar accretion and stellar microlensing are necessary. 
\end{abstract}

\maketitle

\section{Introduction}

        As a light beam travels through the matter distribution in space to reach us
        it gets both sheared and focused. If the distortions in the wavefront are
        large enough, there is going to be multiple images due to strong
        gravitational lensing. This effect was first observed by \cite{Walsh79} for
        the doubly-imaged quasar (Q0957+561). Being bright compact light sources, quasars
        can be observed up to high redshifts and are excellent candidates as light sources
        in strong lensing systems. Many such systems have been
        since observed and are of particular interest in cosmology. They can be both
        used to study the properties of dark matter and the cosmological parameters such
        as the Hubble constant \cite{Courbin02,Suyu17}.
        
        The time delay between different lensed images is the only parameter which depends
        on the cosmological length scales and hence can be used to measure the Hubble parameter.
        This was first proposed by \cite{Refsdal64} and lead to dedicated monitoring programs
        such as Cosmological Monitoring of Gravitational Lenses (COSMOGRAIL). There have been many different approaches to measure the time delay
        from long term brightness measurements of the images in strongly lensed systems. A comparison
        of some of these methods against synthetic data generated by the Time Delay Challenge can be
        found in \cite{Liao15}. 
        
        Another area where strong lensing can be particularly useful is in detecting (or constraining) the distribution and abundance
        of dark matter. For example,  it can be employed to study the amount of dark matter in the
        lens system (e.g. \cite{Bolton08}), or the sizes of its substructures (e.g. \cite{Dalal:2001fq, Hezaveh:2016ltk}).
        
         Less well-known is how time variability of the images can also  be used to measure the statistics of CDM nanostructures (or microhaloes). This is known as 
        the transient weak lensing effect, and is induced by the moving dark matter microhaloes that cross the lines of sight towards
        multiply-imaged quasars \cite{Rahvar14}. In this work, we search for the transient weak
        lensing signal in strongly lensed quasar systems, leading to constraints on the (linear and nonlinear) dark matter power
        spectrum. As a by-product of our analysis we also measure the strong lensing time delay, as well as the temporal power spectrum of quasar accretion flow.
        
        The paper is structured as follows: We describe the data used in this work in Section
        \ref{sec:data}. Next, our method is described in detail in section \ref{sec:method} and
        the details of the parameter estimation techniques are discussed in Section
        \ref{sec:sampling}.
        The limitations arising from the finite size of the light emitting region in source quasars
        is discussed in section \ref{sec:finite}.
        Section \ref{sec:test} presents the results of applying our method to the Time Delay Challenge
        data.
        The results for internal and lensing power spectra, as well as constraints on the $\Lambda$CDM linear spectrum are presented in Section \ref{sec:results}, which are  followed by the Conclusions.

\section{Datasets}
\label{sec:data}
      
      Strongly lensed quasar systems have been monitored by several different groups.
      This includes radio observations such as \cite{Rumbaugh15} and optical measurements.
      Optical measurements include dedicated campaigns such as COSMOGRAIL as well as
      observations by other groups such as the OGLE gravitational microlensing
      group \cite{Kochanek04}.
      We use the data made publicly available by COSMOGRAIL to demonstate our method.
      
      COSMOGRAIL is a project aimed at
      constraining the cosmological parameters by monitoring strongly lensed quasars.
      It has monitored the light curves of a few well known lensed quasars over the
      course of a decade in an attempt to 
      measure the time delay between different images.
      The data consists of R-band light curves for each lensed image.
      There are six publicly available 
      lensed systems, namely HE 0435−1223, SDSS J1001+5027,
      RX J1131−1231, SDSS J1206+4332, HS 2209+1914 and DES J0408-5354 \cite{cosmograilweb}.\\
      To validate our method we have used the Time Delay Challenge (TDC) dataset.
      It includes thousands of light curves that are made to represent different
      data quality and observational strategies, as well as
      many realistic features present in real data such as periods of
      missing data and the effect of gravitational microlensing by stars in the
      lensed galaxy. As such, the TDC dataset serve as an independent test to measure the
      performance of our method.
      
\section{Method description}
\label{sec:method}
      The $\Lambda$CDM model predicts a hierarchy of dark matter haloes on different length scales. While baryonic
      matter can cool and form galaxies in the potential wells of larger dark matter haloes,
      the cooling time is too long in the smaller halos, which are then non-luminous. 
      As the light coming from a distant quasar travels towards the observer it encounters
      several dark matter haloes of various sizes, each one inducing an additional weak
      lensing effect. Since the haloes are moving across the line of sight, the lensing effect is time variable.
      In \cite{Rahvar14} a relation between the dimensionless matter
      power spectrum and the temporal power spectrum of the lensing amplification was derived. 
      In this section, we describe how the temporal lensing power spectrum and the time delay
      are constrained by strongly lensed quasar light curves. Section \ref{sec:finite} describes
      how the constraints on temporal lensing power spectrum are related to the limits on the 
      dark matter power spectrum.
      
      In the following, we describe how the likelihood function for a set of observations depends on the free parameters of the model. The data is the observed magnitude of the quasar images
      in a strongly-lensed system over a period of time. The model consists of time delays
      between different images and two power spectra: the temporal power spectrum
      of weak lensing amplification and the temporal power spectrum of intrinsic quasar
      magnitude variations.

      A given image in the multiply-imaged quasar system is labeled with the subscript $a$.
      Each measurement of the apparent magnitude for image $a$ is decomposed into three parts,  
      an intrinsic part, $m(t_{ia}+T_{a})$, a part caused by the
      gravitational lensing effect, $L_{a}(t_{ia})$, and the  
      measurement error, $n_{ia}$:
      \begin{equation}
        I_{ia}=m(t_{ia}+T_{a})+L_{a}(t_{ia})+n_{ia}
        \label{eq:lightcurve1}
      \end{equation}
      where $t_{ia}$ is the time at which the $i$th measurement is done for image $a$.
      $T_{a}$ represents the gravitational time delay for image labeled
      by $a $ (with $T_1=0$).
      The vector $I$ contains the light curve data for lensed images of the same quasar
      stacked together. 

      The subscript $i$ runs over different time steps from $1$ up to the total number of time
      steps $N_T$.
      The subscript $a$ takes $N_{I}$ different values where $N_{I}$ is the number of images. 
      These two indices can be combined into a single index (represented by greek letters)
      and defined as $\mu = i + N_{T}\times(a-1)$.
      Using this convention, Equation \ref{eq:lightcurve1} can be re-written as:
      \begin{equation}
        I_{\mu}=m(t_{\mu}+T_{a})+L_{a}(t_{\mu})+n_{\mu}.
        \label{eq:lightcurve2}
      \end{equation}  
      The covariance matrix for apparent magnitude measurements can be expanded in terms of power
      spectra:
      \begin{align}
        C_{\mu\nu} &= \langle I_{\mu}I_{\nu}\rangle = \delta_{\mu\nu}\sigma^{2}_{\mu}+ \nonumber \\ 
        & \int \frac{\rm{d}\omega}{2\pi}e^{i\omega(t_{\mu}-t_{\nu})} [\delta_{ab}P_{L}(\omega)+e^{i\omega(T_{a}-T_{b})}P_{m}(\omega)].
        \label{eq:lightcurve3}
      \end{align}  
      where $P_m(\omega)$ is the intrinsic temporal power spectrum and $P_{L}(\omega)$
      is the lensing temporal power spectrum. Note that we have assumed that the lensing effect is uncorrelated across different images, which is the key property we use to distinguish the intrinsic from the lensing temporal power spectra. 
      
      Now we divide the relevant part of frequency space into  $N_{F}$ frequency bins
      and approximate the power spectra using the Heaviside step functions as:
      \begin{align}
        &P_{L}(\omega)=\sum^{N_{F}}_{l = 1}p_{l}K_{l} (\omega),\nonumber \\
        &P_{m}(\omega)=\sum^{N_{F}}_{l = 1}p_{l+N_{F}} K_{l}(\omega),\nonumber \\
        &K_{l}(\omega)=\Theta(\omega-\omega_l+\Delta\omega_l)\Theta(\omega_l+\Delta\omega_l-\omega).
        \label{eq:lightcurve4}
      \end{align}  

      We assume that $P_{L}(\omega)$ and $P_{m}(\omega)$ are even functions while
      $p_{l}$'s are unknown weights to be estimated using the data.
      The covariance matrix can then be rewritten as:
      \begin{equation}
        C_{\mu\nu}=\delta_{\mu\nu}\sigma_{\mu}^{2}+\sum^{2N_{F}}_{l=1}p_l\widetilde{K}^{l}_{\mu\nu}
        \label{eq:lightcurve5}
      \end{equation}  
      Where $\widetilde{K}^{l}_{\mu\nu}$ for $1\leq l \leq N_{F}$ is defined as:
      \begin{align}
        &\widetilde{K}^{l}_{\mu\nu} =\delta_{ab} F^{l}(t_{\mu}-t_{\nu})\nonumber \\
        &\widetilde{K}^{l+N_{F}}_{\mu\nu} =F^{l}(t_{\mu}-t_{\nu}+T_{a}-T_{b}) \nonumber \\
        &F^{l}(\Delta t)= \int \frac{\mathrm{d}\omega}{2\pi}e^{i\omega\Delta t}K^{l}(|\omega|)
        \label{eq:lightcurve6}
      \end{align}
      where $\Delta T_{ab} = T_{a}-T_{b}$ is the time delay between different images.
      For each unknown parameter set $p_{l}$ and $\Delta T_{ab}$, the  chi-squared can be written in terms 
      of the data vector $I$ and the covariance matrix $C$ as:
      \begin{equation}
        \chi^2(p_l, \Delta T_{ab})=I^{t}CI+\log{\left[\mathrm{det}(C)\right]}\label{eq:lightcurve7}
      \end{equation}

      Having the likelihood function $\equiv \exp(-\chi^2/2)$, we use a Markov Chain Monte Carlo (MCMC) to explore
      the parameter space and find the best fit values together with their uncertainties.

      The power spectra functions are considered between a minimum and a maximum
      frequency corresponding to a minimum time scale $T_{min}$ equal to one third of the median
      of the time difference between data points and a maximum time scale $T_{max}$
      which is equal to three times the time difference between the first and the last data points. $N_{F}$ has 
      been set to $9$ while the width of frequency bins has been chosen such that they are equality 
      spaced in logarithmic scale. In addition there is a bin at very low frequency to take out the
      very long scale variations.

\section{MCMC Sampling}      
\label{sec:sampling}

      In this section we describe the method used to explore the model parameter space
      and find their posterior probability density function.

      First, we need to define a likelihood function and choose priors on model parameters.
      In this work, we choose the following form for the likelihood function:
      \begin{equation}
        \mathcal{L}(p_{l}, T_a) = \exp{\left(-\frac{\chi^2} {2}\right)}
      \end{equation}
      We further choose a flat prior over a reasonably wide range for all the parameters in the model.
      The details on the ranges are presented with the results below.

      Having a large number of parameters, MCMC methods would be a natural choice. We tried a range
      of MCMC algorithms such as Metropolis-Hastings and Gibbs sampling with adaptive step size tuning
      but they generally struggled to yield reliable answers and suffered from convergence issues.
      A combination of Affine-Invariant MCMC \cite{goodman10} and parallel tempering proved to give reliable estimates.
      Here is a brief description of the algorithm used.
      
      Affine invariant MCMC is a particular form of ensemble sampling that performs equally well on a parameter space
      mapped by any Affine transformation. In particular it can sample highly skewed distributions with linear correlations very efficiently.
      It's also straightforward to parallelize and hence take advantage of the available high performance computing facilities. These methods only  have a few
      hyperparameters and can be efficiently used on a large number of problems with minimal need for tuning.
      Having a highly irregular and spiky likelihood surface, the Affine-invariant ensemble sampler would
      spend a long time in local extrema and would suffer from slow convergence. To circumvent this problem the
      ensemble sampler was combined with a parallel tempering scheme \cite{Earl05}.

      Parallel tempering makes many copies of the likelihood function modified by a ``temperature'' parameter:
      \begin{equation}
        \mathcal{L} \propto \exp{\left(-\frac{\chi^2} {2T}\right)},
      \end{equation}
      where $ 1 \leq T \leq T_{max}$ is the temperature parameter. At $T = 1$ we have the original likelihood that
      we wish to sample. We run an independent ensemble sampler at each temperature and let the chains at
      different temperatures swap their positions in the parameter space after many Monte Carlo steps. This happens
      with a probability given by:
      \begin{equation}
        \min{\left(1, \frac{\mathcal{L}(\vec{x}_2,T_2)}{\mathcal{L}(\vec{x}_1, T_1)}^{1/T_1 - 1/T_2}\right)},
      \end{equation}
      where $T_2 > T_1$. $\vec{x}_1$ and $\vec{x}_2$ are the positions of the two chains in the parameter space.  In this way the high-temperature chains easily move in the parameter space
      and visit places that would have been difficult for the low-temperature physical chains to visit. By performing
      position swaps, the physical chains can sample the region allowed by the priors effectively even
      for hard to sample multimodal distributions. The posterior probability distribution function is then
      given by the density of the lowest temperature chains ($T = 1$)  only.
      
      The choice of temperatures has an important effect on the performance of the sampler. Firstly, the maximum
      temperature should be high enough to allow the chains to effectively move everywhere within the region
      permitted by the priors. Secondly, the temperature difference between adjacent temperatures should be small
      enough to allow position swaps to happen often. The choice of temperature ladder is not clear {\it a priori}.
      We used a method to adaptively tune the temperatures so that we get uniform swapping acceptance rate
      between the adjacent temperatures. This avoids having a bottleneck in propagation of positions visited in the
      parameter space by the highest temperature chain to the lowest temperature physical chains \cite{vousden16}.
      The parallel tempering method could also be trivially parallelized which is very important in our case
      since likelihood calculations are computationally expensive and the runtimes can be otherwise very long.

\section{Finite Size Effect}
      \label{sec:finite}
            
      Assuming the quasar is a point source, the lensing temporal power spectrum can be
      calculated and  is given by \cite{Rahvar14}:

      \begin{align}
        &\omega P_{L}(\omega) = 18\pi^2H_0^4{\Omega^{(0)}_m}^2\int_0^{\chi_s}\left(1-\frac{\chi^{\prime}}{\chi_s}\right){\chi^{\prime}}^2\mathrm{d}\chi^{\prime}\nonumber \\
        &\int_0^{\infty}\mathrm{d}v~e^{-v^2/\sigma^2} \left(\frac{v}{\sigma} \right)^2\frac{\Delta^2}{\omega}(1+z_{\chi^{\prime}})^3,
      \end{align}
      where $\chi_s$ is the comoving distance to the quasar, $z_{\chi}$ is the
      redshift at comoving distance $\chi$, $\sigma$ is the velocity dispersion
      of dark matter halos and $\Delta^2$ is the dimentionless
      matter power spectrum. For this work, we adopt $\sigma \simeq 500~ \rm{km/s}$, which is dominated by the cosmological bulk flows on large scales ($\sim 30$ Mpc) \cite{Rahvar14}.
      
      We will recalculate this to take into account the finite size of the quasars' emitting region and
      generalize this result to include the effect of the finite size
      of the source. 
      Assuming a radial surface brightness profile given by a function $f(r)$ 
      we find the following formula for the lensing temporal power spectrum:

      \begin{align}
        &\omega P_{L}(\omega) = 18\pi^2H_0^4{\Omega^{(0)}_{m}}^2\int_0^{\chi_s}\left(1-\frac{\chi^{\prime}}{\chi_s}\right)^2\chi^{\prime 2}\mathrm{d}\chi^{\prime}\nonumber \\
        &\int_0^{\infty}\frac{\mathrm{d}v}{\sigma^2}e^{-v^2/2\sigma^2}
        \int_{\omega/v(1+z)}^{\infty}\frac{\mathrm{d}k_{\perp}}{2\pi} \frac{\omega(1+z_{\chi^{\prime}})\Delta^2F^2_0(k_{\perp})}{k^2_{\perp}\sqrt{k_{\perp}^2-\omega^2/v^2(1+z_{\chi^{\prime}})^2}},
      \end{align}
      $K_{\perp}$ is the transverse wavenumber and $F_0(K_{\perp})$ is the normalized Hankel transform of surface brightness
      $f(r)$ which is given by:
      \begin{equation}
        F_0(k_{\perp}) = \frac{\int_0^{\infty}f(r)J_0[rk_{\perp}(1+z)]r\mathrm{d}r}{\int_0^{\infty}f(r)r\mathrm{d}r}.
      \end{equation}

\begin{figure}
        \includegraphics[scale = 0.23]{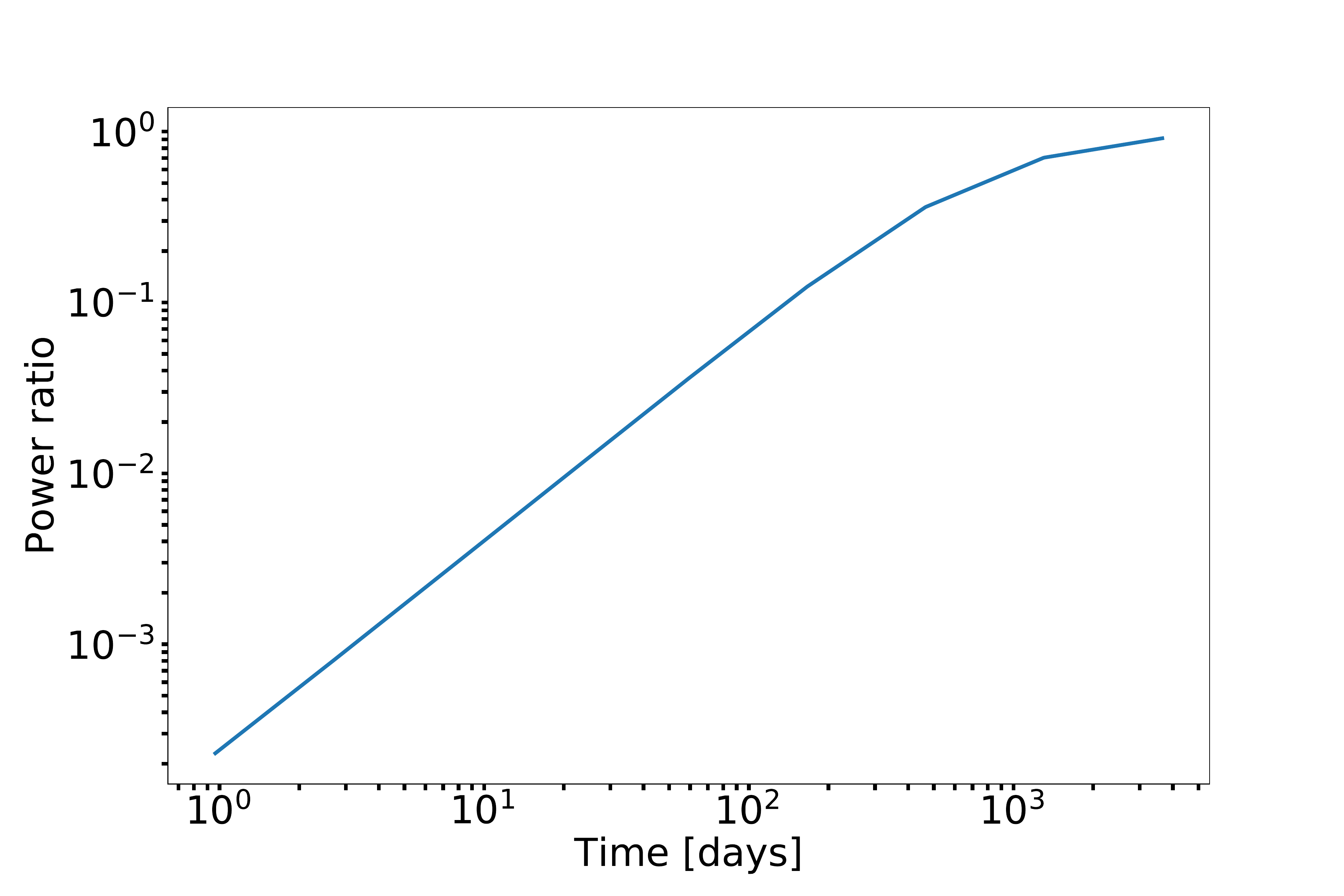}
        \caption{Ratio of temporal power spectrum for an extended source of
          size $0.1 ~ \rm{light-day}$ to that of a point source. The horizontal axis is $T \equiv 2\pi/\omega$.  
      }
        \label{powerRatio}
      \end{figure}

      Figure \ref{powerRatio} shows the ratio of the lensing temporal power spectrum
      for an extended source of size 
      $0.1 ~ \mathrm{light-day}$
      compared to a point source quasar. The extended source is assumed to be a
      Shakura-Sunyaev disk \cite{Shakura73} radiating as a black body. As can be seen here the finite size
      of the source can drastically suppress the power on short time scales.  



      Another factor to consider is the effect of strong lensing on amplifying the fluctuations from the
      transient weak lensing \cite{Rahvar14}. The effect is an enhancement in the power by a factor of:
      \begin{equation}
        \left(\frac{2(\kappa^2+\gamma^2)^{1/2}}{|(1-\kappa)^2-\gamma^2|}\right)^2,
      \end{equation}
      where $\kappa$ and $\gamma$ are the convergence and shear.
      \comment{
      Since the
      values for $\kappa$ and  $\gamma$ were generally unknown we choose a typical value
      of $20$ for this factor from here on. In each individual case depending on whether
      this factor is larger or smaller the inferred values for the power spectrum may
      move by an order of magnitude. }

      The values of $\kappa$ and $\gamma$ are generally not known for the source quasars
      so the enhancement cannot be calculated. If $\gamma^2 \ll 1$ the enhancement factor
      can be approximated as
      \begin{equation}
4(-1+\sqrt{\mu})^2\mu, 
\end{equation}
      where $\mu = 1/|(1-\kappa)^2 - \gamma^2|$ is the lensing magnification \cite{Takahashi11}. We used the estimated magnifications, calculated through
      lens modelling for three quasars, namely, HE-0435−1223 \cite{Wisotzki02}, RX-J1131−1231 \cite{Blackburne10}
      and DES-J0408-5354 \cite{Agnello17} to find the enhancement factor.
      For all the other quasars we used the average enhancement
for the known quasars which is $\approx 400$.

\section{Recovering time delays and the time-delay challenge}
\label{sec:test}


      As discussed in Section \ref{sec:method}, our model consists of the intrinsic
      and lensing power spectra and the time delay between the light-curves. Therefore,
      we can recover the time delay for the quasar images. In this section, we test
      the ability of our pipeline to recover the correct time delays.
      
\begin{figure}
        \includegraphics[width = \columnwidth]{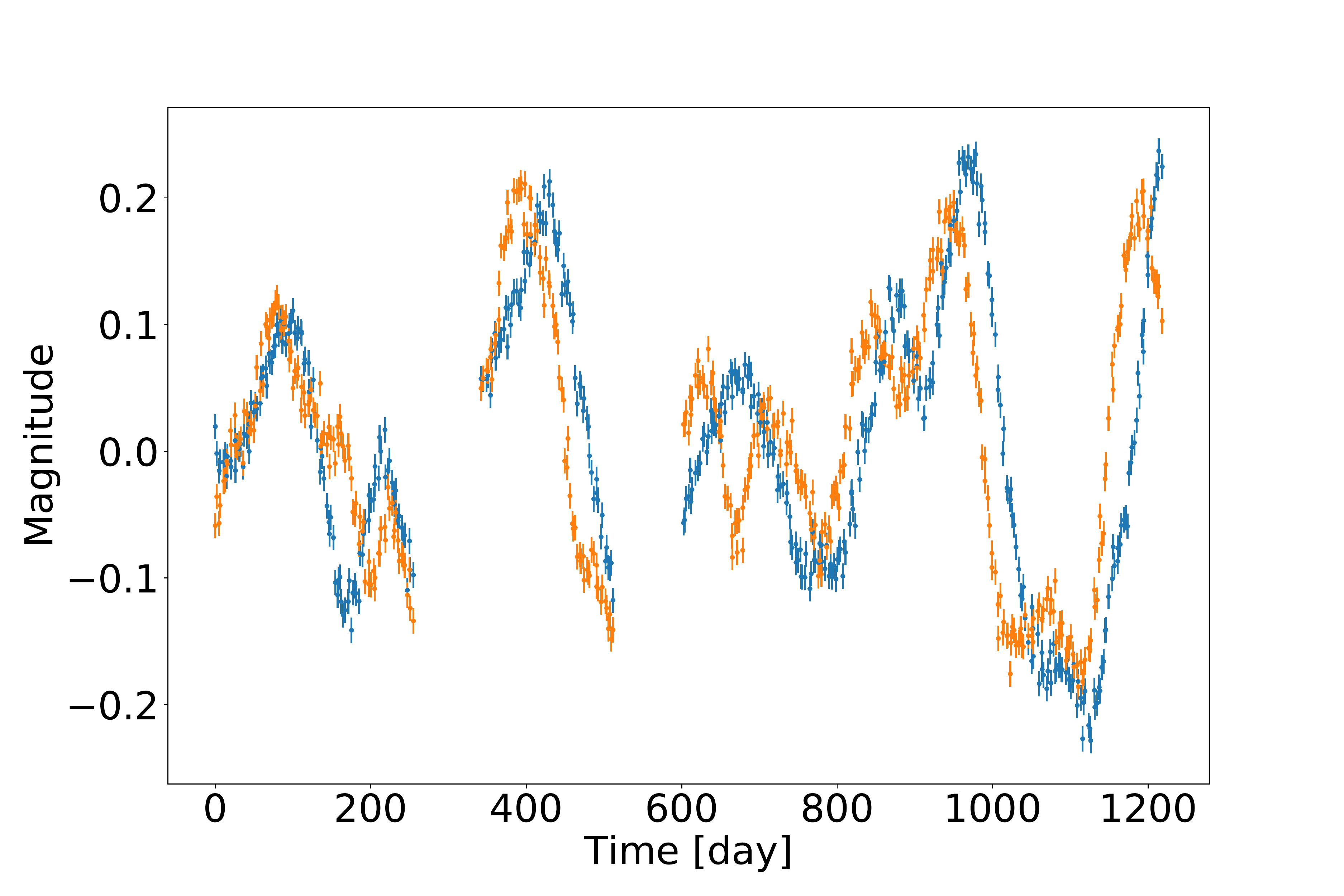}
        \caption{Synthetic light curves for the two quasar images.}
        \label{fig:MockLightCurve}
      \end{figure}

      The first test involves generating synthetic light curves and using them in our 
      pipeline. The lightcurves are generated using predefined lensing and intrinsic 
      power spectra and time delays. We then compare the recovered values to 
      the actual input values. The data is generated to mimic the observational 
      strategy adopted by COSMOGRAIL. The time sampling is randomized and the time shift 
      due to strong gravitational lensing is included. 
      The light curves include observational errors and missing
      data intervals corresponding to non-observing seasons. Figure \ref{fig:MockLightCurve}
      shows an example of such a light curve for two images of a strongly lensed quasar. 

\begin{figure}
        \includegraphics[width = \columnwidth]{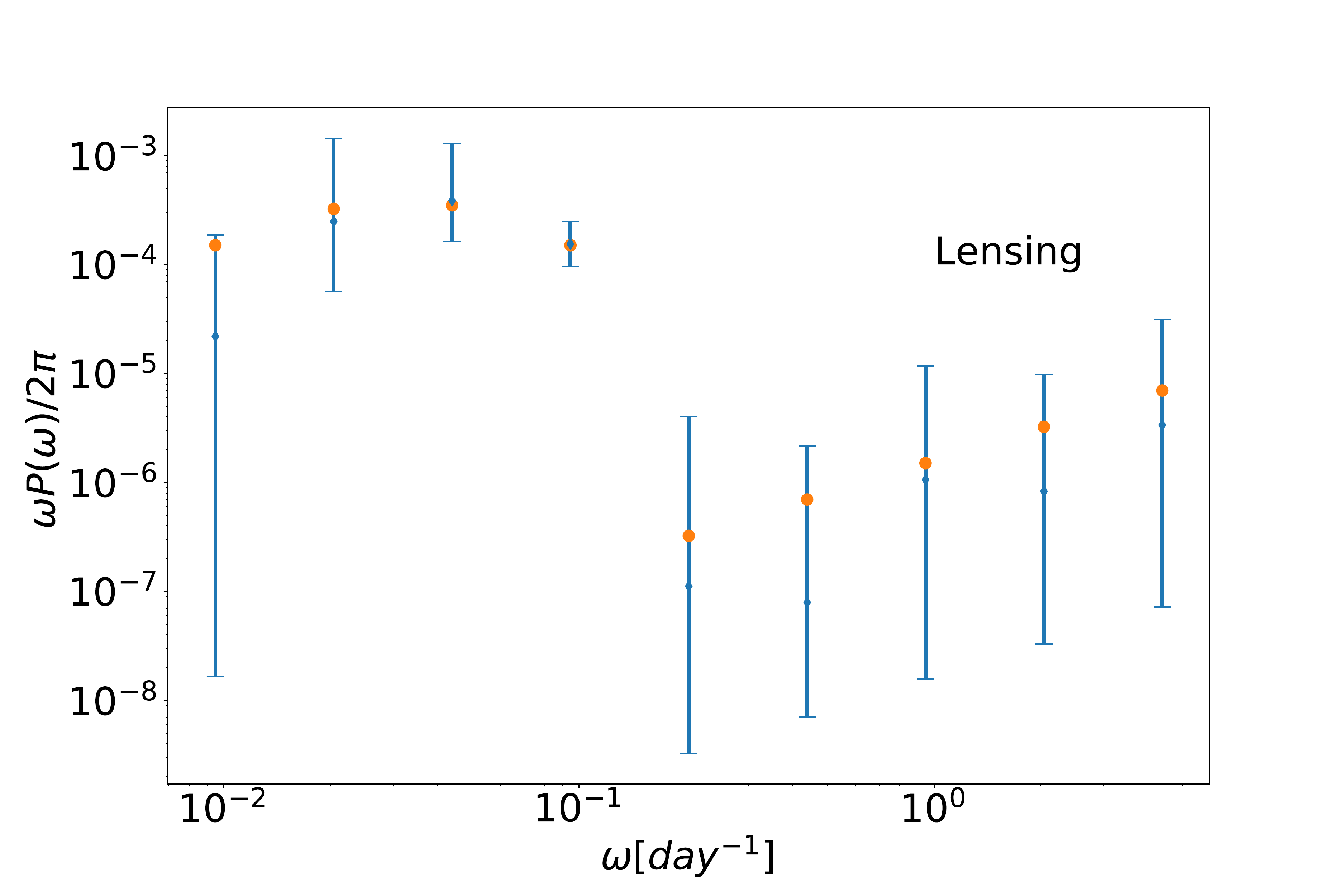}
        \caption{Lensing power spectrum for the synthetic quasar light curve measured by
          our method.
          The orange dots are the actual values used to generate the light curve.
        The errorbars show the three sigma uncertainty region.}
        \label{fig:MockLensing}
      \end{figure}

\begin{figure}
        \includegraphics[width = \columnwidth]{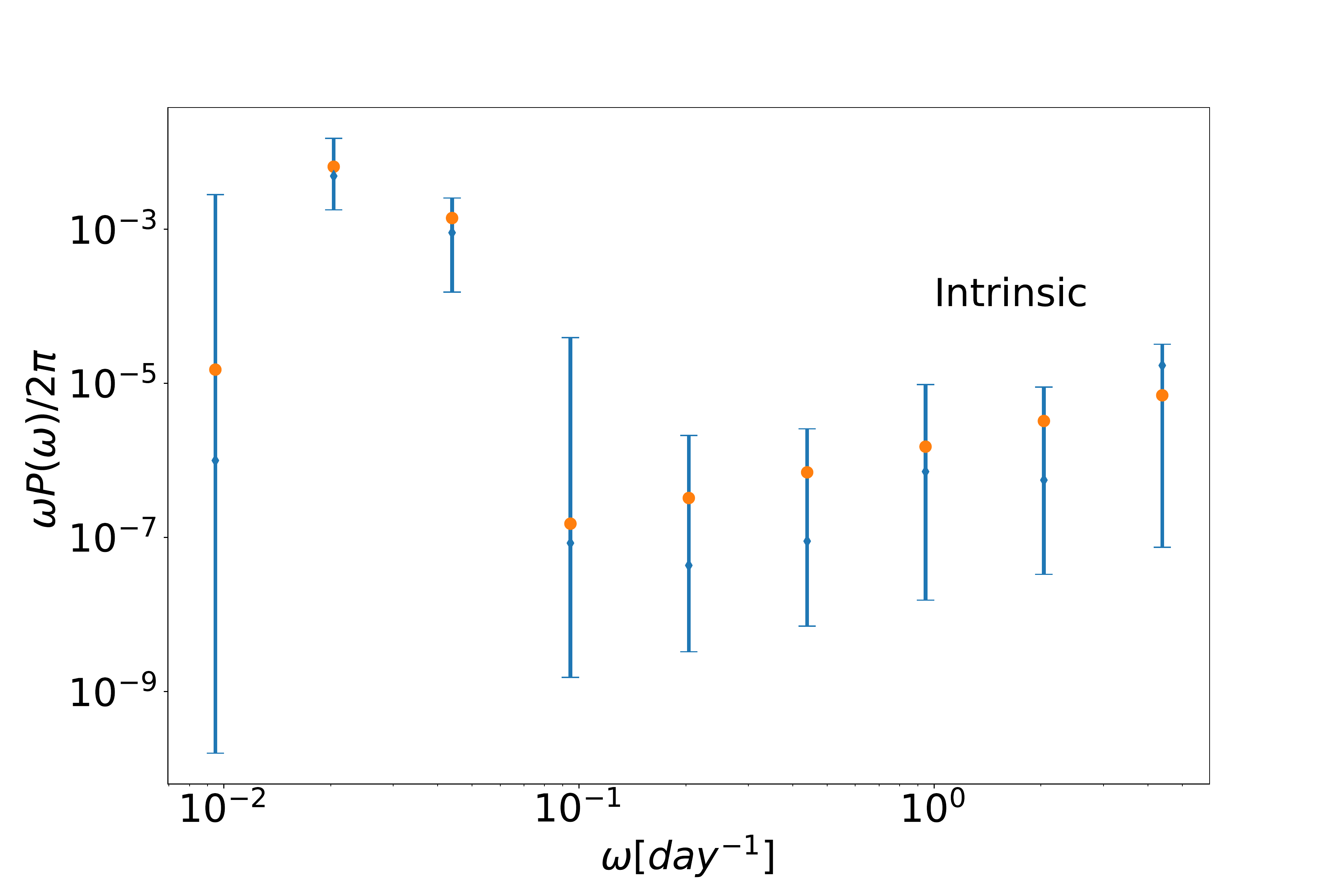}
        \caption{Intrinsic power spectrum for the synthetic quasar light curve measured by
          our method.
          the orange dots are the actual values used to generate the light curve.
        The errorbars show the three sigma uncertainty region.}
        \label{fig:MockIntrinsic}
      \end{figure}

      For each run several tests are performed to ensure the MCMC chains have converged.
      For this example light curve the true time delay $\Delta T = 23 ~\rm{days}$ was recovered as
      $\Delta T = 25.1 \pm 3.7 ~\rm{days}$. In addition the lensing and intrinsic power
      spectra were recovered as shown in Figures \ref{fig:MockLensing} and \ref{fig:MockIntrinsic}.

      The next test involves using TDC light curves \cite{Liao15}. There are thousands 
      of generated light curves separated into different classes with different data quality 
      and observational strategies. These are called different rungs and the differences  
      include different cadence, total observational timespan and dispersion in the cadence. 
      We chose light curves from all 5 available rungs and compared the recovered time delays 
      to the true values. Table \ref{tbl:dtTest} shows the results of recovered time delays 
      for randomly selected  TDC light curves. 

      \begin{table}
       \begin{tabular}{cccc}
            
        \hline
        \multicolumn{4}{r}{Time delay[days]} \\
        \cline{3-4}
        rung    & $N_I \quad$ & Actual & Recovered \\
        \hline
        0      & 2 & $5.78\quad$    & $5.34^{+0.60}_{-0.54}$      \\
        1      & 2 & $14.23\quad$    & $14.06^{+0.09}_{-0.08}$      \\
        2      & 2 & $28.44\quad$        & $29.16^{+0.96}_{-0.71}$       \\
        3      & 2 & $57.53\quad$     & $56.71^{+0.91}_{-1.29}$      \\
        4      & 2 & $27.2\quad$     & $26.91 ^{+0.29}_{-0.29}$      \\
        \hline
      \end{tabular}
      \caption{Recovered values vs actual time delay values for 
        randomly selected TDC light curves. $N_I$ is the number of 
        images and error bars are $1\sigma$ significance level.}
      \label{tbl:dtTest}
    \end{table}


\section{Results} 
\label{sec:results}

      In this section, we discuss the results of applying our method to COSMOGRAIL light 
      curves. The results are divided into three separate sections discussing the time-delays,
      the limits on the dark matter power spectrum and the power spectrum for quasar variability.

Once the limits on the nonlinear dark matter power spectrum are found, one can use the
      stable clustering hypothesis to transform these into limits on the linear dark matter power
      spectrum(\cite{Peacock96, Smith03, Zavala15}). Here, we shall use the particle phase space average density (P$^2$SAD) modelling, provided in \cite{Zavala15}, which is calibrated against numerical N-body simulations \cite{2008MNRAS.391.1685S} The model is inspired by the stable clustering hypothesis in phase space and supported by the remarkable universality of the clustering of dark matter in phase space as measured by P$^2$SAD across simulated haloes of different masses, environments, and redshifts. On small scales and for primordial power spectra which are reasonably similar to $\Lambda$CDM, we can fit the P$^2$SAD predictions for $\Delta^2_{\rm NL}(k_{\rm NL})$  by power laws in terms of the linearly extrapolated power spectrum $\Delta^2_{\rm L}(k_{\rm L})$: 
       \begin{equation}
        \Delta^2_{\rm NL}(k_{\rm NL}) \approx a \left[\Delta^2_{\rm L} (k_{\rm L})\right]^{\alpha}, \quad
        k_{\rm NL} \approx b  \left[\Delta^2_{\rm NL}(k_{\rm NL})\right]^{1/3} \times k_{\rm L},
      \end{equation}
      where $a = 0.24, b = 1.12 ~\rm{and}~ \alpha = 3.05$. 
      
      The inferred linear power spectrum $\Delta^2_{\rm L}(k_{\rm L})$ can then converted to the primordial power spectrum ${\cal P}_{\cal R}(k_{\rm L})$, using the $\Lambda$CDM linear transfer and growth functions. For Planck 2015 cosmology the conversion factor is:
      \begin{equation}
      \Delta^2_{\rm L}(3 ~{\rm pc}^{-1}) = 1.5 \times 10^{11} {\cal P}_{\cal R}(3 ~{\rm pc}^{-1})
      \end{equation}
      
      \subsection{Time delays}

      In this section, we present the recovered time-delay values and compare them to the results obtained by the COSMOGRIAL collaboration. Table \ref{tbl:dt} summarizes the results. It shows the time delay values obtained in this work and the time delay values obtained in previous works. It should be noted that in cases where multiple previous estimates existed only one is quoted in the table. The last column in the table provides the references for the quoted time delay values.The forth column shows the name of the images used for calculating the time delays. The name designations follows that of the corresponding reference given in the last column. There are two sets of results for DESJ0408-5354 in the table. The first result is when only two of the three available light curves were used. The next two lines show the result when all three light curves were fitted simultaneously. 

      \begin{table}
       \begin{tabular}{ccccc}
            
        \hline
        \multicolumn{5}{c}{Time delay[days]} \\
        \cline{2-3}
        Name   & This work & Previous works & Images & Ref. \\
        \hline\\
        HE0435-1223   & $8.56^{+0.05}_{-0.06}$ & $ 8.4 \pm 2.1$ & BA &\cite{Bonvin17}     \\[0.3cm]
         RXJ1131-1231 & $0.45^{+0.05}_{-0.90}$ & $0.7 \pm 1.0$ & BA & \cite{Tewes13} \\[0.3cm]
         HS2209+1914 & $22.11^{+2.95}_{-3.33}$ & $20.0 \pm 5.0$ & BA & \cite{eulaers13} \\[0.3cm]
         J1206+4332 & $109.31^{+2.27}_{-2.34}$ & $111.3 \pm 3.0$ & AB & \cite{eulaers13} \\[0.3cm]
         J1001+5027 & $116.11^{+2.11}_{-2.62}$ & $119.3 \pm 3.3$ & BA & \cite{kumar13} \\[0.3cm]
         DESJ0408-5354 & $113.91^{+39.34}_{-1.46}$ & $112.1 \pm 2.1$ & BA&  \cite{courbin17} \\[0.3cm]
         \multirow{2}{*}{DESJ0408-5354}& \multicolumn{1}{c}{$113.93^{+26.92}_{-11.58}$} & \multicolumn{1}{c}{$112.1 \pm 2.1$} & \multicolumn{1}{c}{BA} & \multicolumn{1}{c}{\cite{courbin17}}\\[0.3cm]
         & \multicolumn{1}{c}{$151.44^{+47.01}_{-17.57}$} & \multicolumn{1}{c}{$155.5 \pm 12.8$} & \multicolumn{1}{c}{DA} & \multicolumn{1}{c}{\cite{courbin17}}\\
         
        \hline
      \end{tabular}
      \caption{Time delay values obtained using our method compared to the previous works. The first column is the name of the quasar system. The second and third column present the values found in this work and the values found in previous works. The fourth column lists the images used to calculate the time delays. The image designations follow the conventions in the corresponding reference given in the last column.}
      \label{tbl:dt}
    \end{table}

      \subsection{Quasar temporal power spectrum}
\begin{figure}[H]
        \includegraphics[width = \columnwidth]{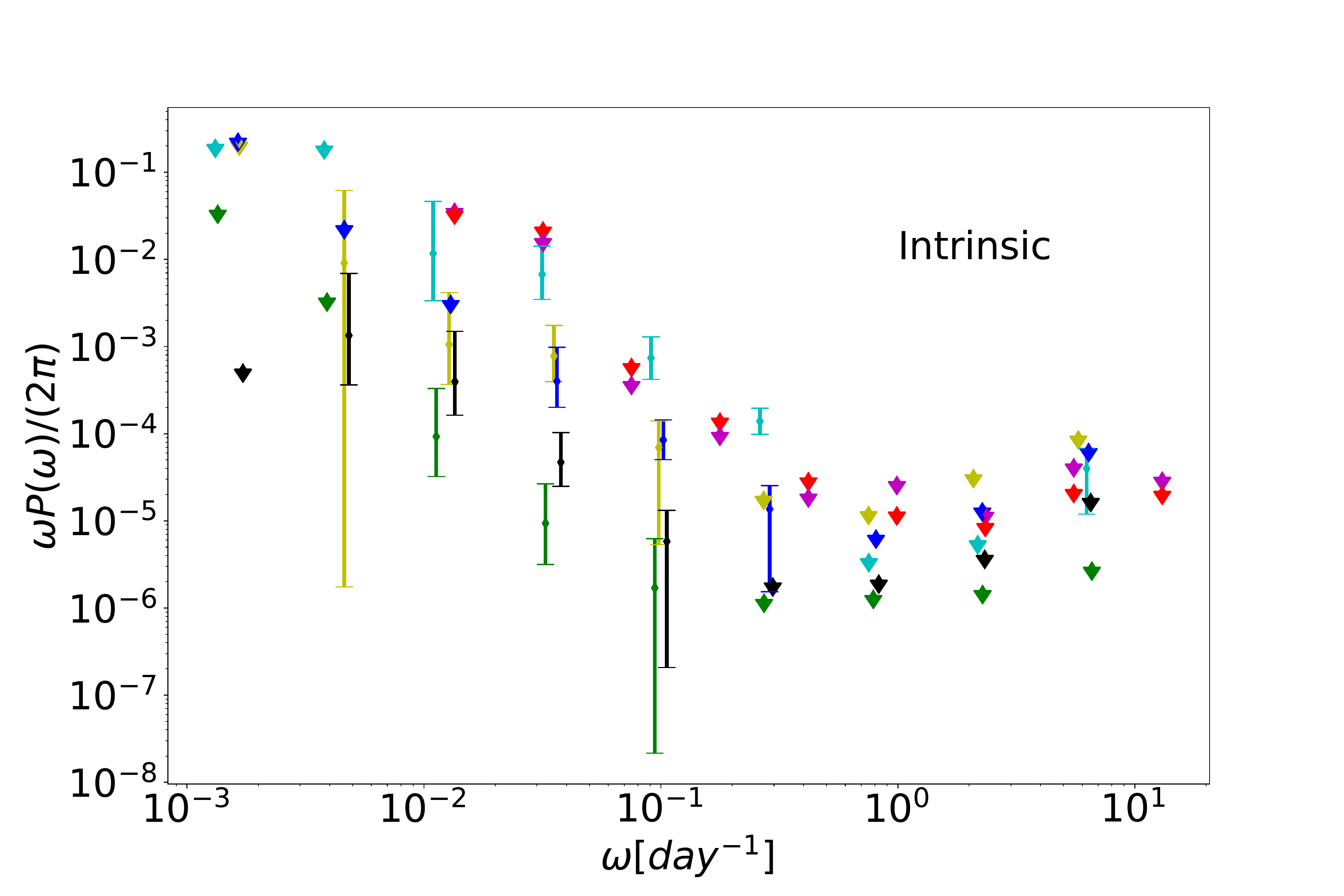}
        \caption{Measurements of the temporal power spectrum for quasar intrinsic variability.
          Error bars are the two sigma confidence intervals.
          Different colors represent different quasars.  The color red is DESJ0408-5354 using two light curves, magenta is the same quasar using three light curves, black is J1001+5027, yellow is J1206+4332, green is HS2209+1914, blue is HE0435-1223 and cyan is RXJ1131-1231.}
        \label{fig:Int_Power}
      \end{figure}

      One of the output products of our pipeline is the temporal power spectrum of intrinsic quasar
      variability. Figure \ref{fig:Int_Power} shows this measurement for the COSMOGRAIL quasars.      
      The low frequency break in the power spectrum can be used to constrain the size of the accretion
      disk (\cite{Lyubarskii97}, \cite{Hawkins07}). The detailed analysis of disk size will be the
      subject of a separate paper.

      \subsection{Dark matter power spectrum}
\begin{figure}[H]
        \includegraphics[width = \columnwidth]{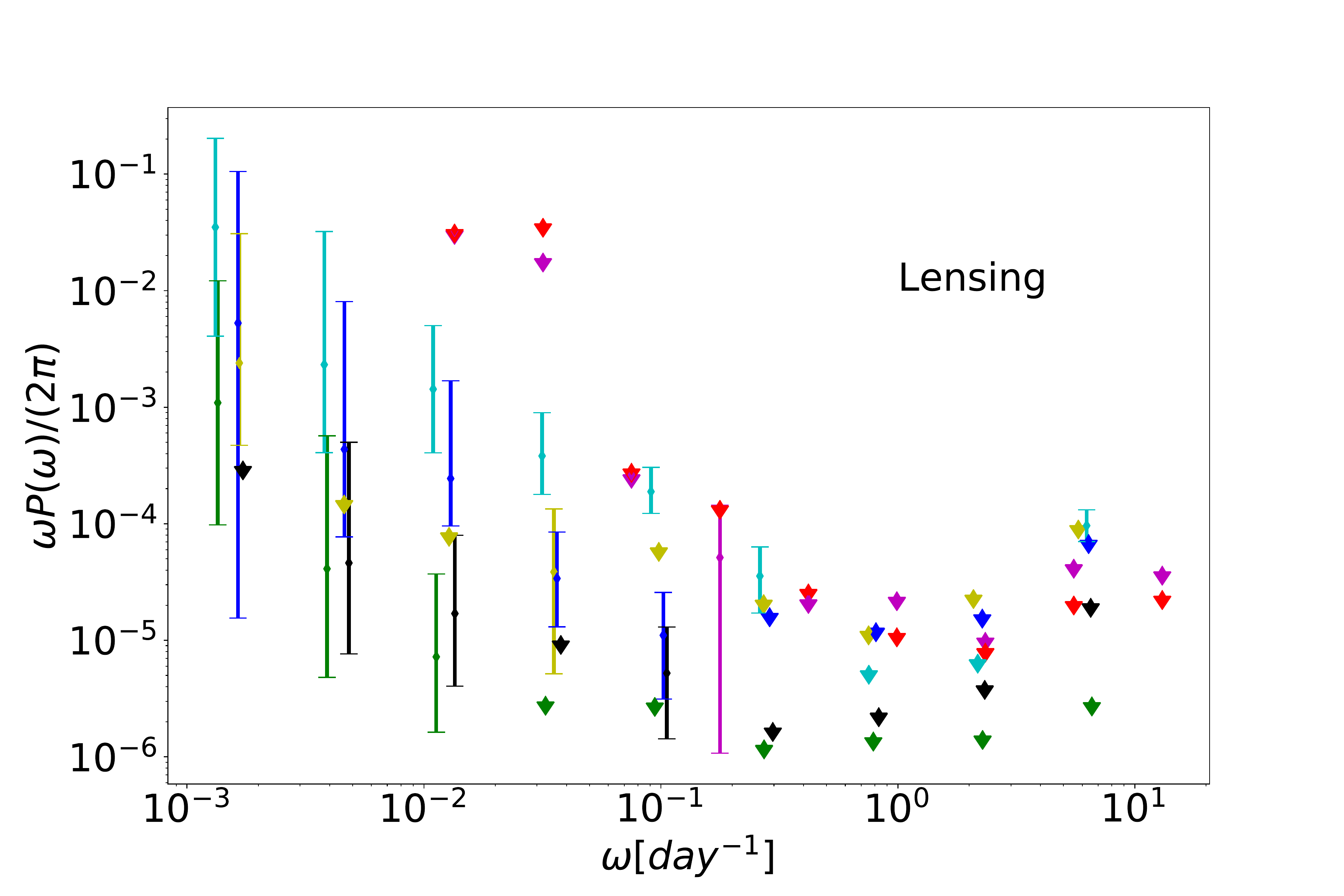}
        \caption{Measurements of the lensing power spectrum. Error
          bars are the two sigma confidence intervals.
          Different colors represent different quasars. The color red is DESJ0408-5354 using two light curves, magenta is the same quasar using three light curves, black is J1001+5027, yellow is J1206+4332, green is HS2209+1914, blue is HE0435-1223 and cyan is RXJ1131-1231.}
        \label{fig:Lns_Power}
      \end{figure}

\begin{figure}
        \includegraphics[width = \columnwidth]{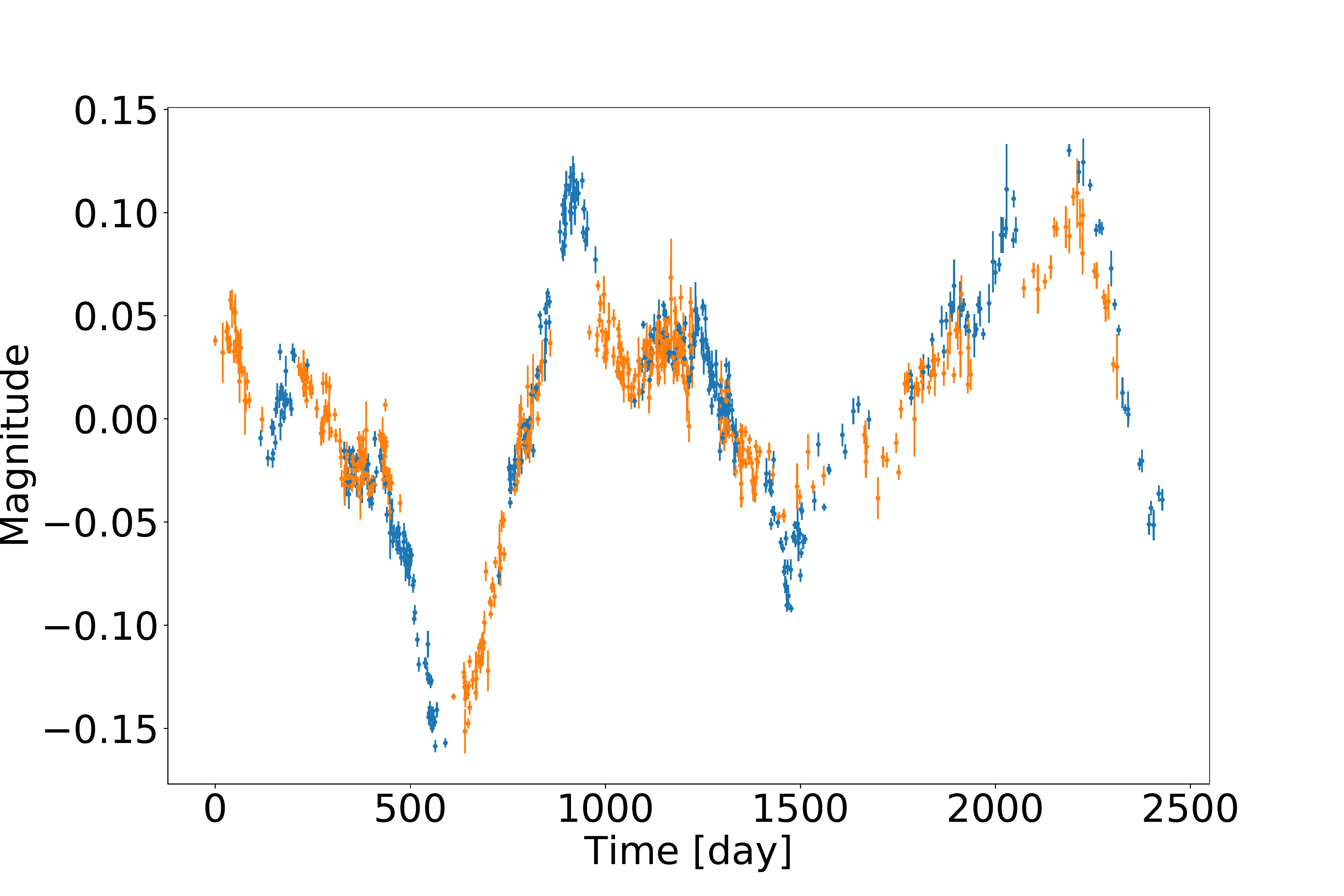}
        \caption{The mean subtracted light curves for the two lensed images in doubly lensed Quasar system J1001+5027. The light curves are shifted by the best fit time delay value. }
        \label{fig:J1001_light_curve}
      \end{figure}

\begin{figure*}[]
        \centering
        \includegraphics[width = \textwidth]{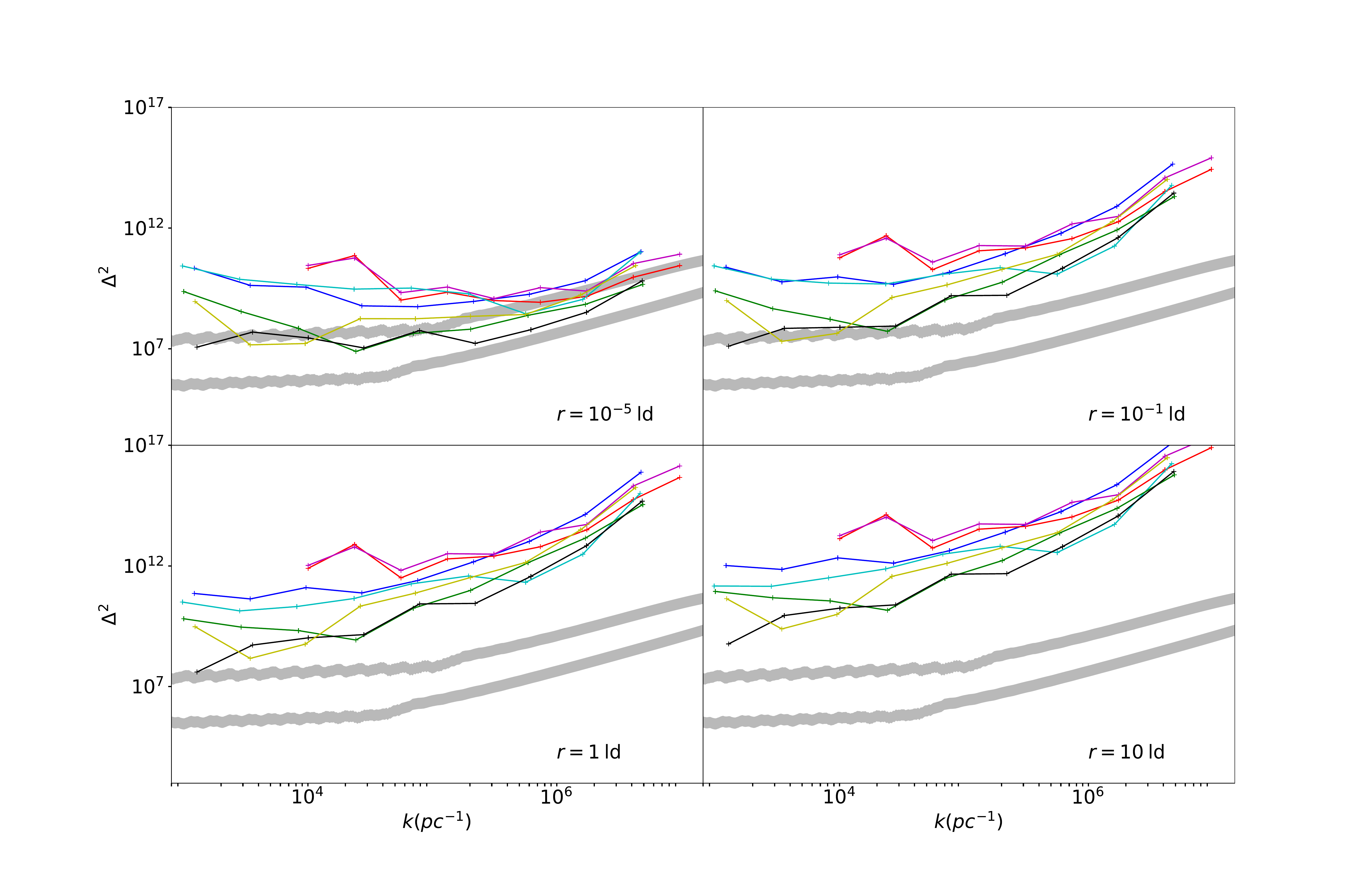}
        \caption{Upper limits on dimentionless dark matter power spectrum at three sigma level. Different sub-plots correspond to different quasar disk sizes given in light days. The lower (upper) grey band shows the prediction for the nonlinear power spectrum using the P$^2$SAD modelling of \cite{Zavala15}, assuming $\Lambda\mathrm{CDM}$ ($5 \times \Lambda\mathrm{CDM}$) linear power spectrum (the band thickness shows 0.5 dex
uncertainty in the substructure volume fraction $0.1 \lesssim f_{\rm subs} \lesssim 0.4$).
          Different colors represent different quasars. The color red is DESJ0408-5354 using two light curves, magenta is the same quasar using three light curves, black is J1001+5027, yellow is J1206+4332, green is HS2209+1914, blue is HE0435-1223 and cyan is RXJ1131-1231. }
        \label{fig:Power_multi}
      \end{figure*}

      The last output from our analysis is the lensing power spectrum. Figure \ref{fig:Lns_Power} shows the results for COSMOGRAIL quasars. 
      The resulting lensing power spectra are then converted into dimensionless matter power spectra. 
      It should be noted that the lensing power recovered by our pipeline contains both the transient weak lensing signal from dark matter
      structures and the gravitational microlensing signal from the stars within the lens galaxy and thus should be interpreted as upper limits.
      These limits depend upon the size of the quasar disk via the finite size effect. We report four sets of
      limits assuming different disk sizes. The
      first set is assuming the light emitting region in the quasar disk size is almost a point source at $10^{-5} ~\mathrm{light-day}$. This may represent the situation
      where most of the light comes from a compact hot spot on the accretion disk.
      The size of the light emitting region in the quasar disk is estimated to be in the range of $\approx 0.1-10 ~\mathrm{light-day}$. We plotted four sets of
      limits for disk sizes of $10^{-5}$, $0.1$, $1$ and $10$ $~\mathrm{light-day}$.
      Figure \ref{fig:Power_multi} shows these limits relative to the $\Lambda\mathrm{CDM}$ predictions using the P$^2$SAD modelling discussed above \cite{Zavala15}.
      The best upper limits are given by the doubly lensed Quasar system J1206+4332 (Black lines in figure \ref{fig:Power_multi}).
      Figure \ref{fig:J1001_light_curve} shows the light curves for the two images in this lensed system.
      
\section{Conclusion and Future Prospects}

    We presented a novel method to simultaneously fit for the time delays of strongly lensed quasars, as well as the power spectra of their intrinsic variability and
    the temporal power spectrum of the gravitational lensing, caused by stellar microlensing and dark matter haloes.
    The recovered time delays are consistent with the previous methods and, depending on the light curve quality, can
    even yield sub percent level accuracy.
    
    We have presented upper limits on the dimensionless dark matter power spectrum over the $10^{-7} - 10^{-3} \mathrm{pc}$ scales which
    remain consistent with the $\Lambda$CDM predictions, despite the dependence on the size the emission region in quasar accretion disks.
    Our strongest limit on the (non-linear) matter power spectrum  is on milliparsec scales, and is given by $\Delta_{\rm NL}^2< 4 \times 10^7$ for $k_{\rm NL} \sim  10^3 {\rm pc}^{-1}$. Using the P$^2$SAD model for nonlinear clustering of CDM nanostructure \cite{Zavala15}, which is calibrated against high resolution N-body simulations, we can translate this limit to a limit on the linear power spectrum. The corresponding constraint on the primordial (linear) scalar power spectrum is given by ${\cal P}_{\cal R} < 3 \times 10^{-9}$ on $k_L \sim$ 3 {\rm pc}$^{-1}$, which is the strongest constraint on these scales, and is within an order of magnitude of the $\Lambda$CDM prediction, assuming a power law power spectrum down to these scales. Finally we were able to measure the power spectrum for the intrinsic quasar variability which could help us study the
    nature of quasar variability and accretion processes.
    
    Future cadence optical imaging surveys, most notably the Large Synoptic Survey Telescope (LSST), are expected to improve the size of the sample of strongly lensed quasars by $\sim$ 3 orders of magnitude, dramatically reducing our statistical errors \cite{2010MNRAS.405.2579O}. However, it is clear that further theoretical modelling in the structure of the emission region in quasar accretion disks, as well as a clean separation of microlensing and transient weak lensing effects (e.g., via the gaussianity of the noise \cite{Rahvar14}) are necessary to lower the upper limits and/or turn them into a detection.

\section{Acknowledgement}
The authors would like to thank Sohrab Rahvar, Alireza Hojjati, Neal Dalal and
Avery Broderick for the useful discussions and comments which helped to
improve this work. JZ Acknowledges support by a Grant of Excellence from the Icelandic
Research fund (grant number 173929-051).
Research at Perimeter Institute is supported
by the Government of Canada through Industry Canada and by
the Province of Ontario through the Ministry of Research and
Innovation.

  


  

\bibliographystyle{unsrt}
\bibliography{Bibliography}
\appendix

\end{document}